\documentclass[%
 reprint,
 amsmath,amssymb,
 aps,
]{revtex4-2}

\usepackage{graphicx}% Include figure files
\usepackage{dcolumn}% Align table columns on decimal point
\usepackage{bm}% bold math
\def\e{\begin{equation}}
\def\f{\end{equation}}
\def\_#1{{\bf #1}}

\def\.{\cdot}

\newcommand*{\affaddr}[1]{#1}
\newcommand*{\affmark}[1][*]{\textsuperscript{#1}}
\usepackage{mathtools}  
\begin{document}

\preprint{APS/123-QED}

\title{Engineering Equifrequency Contours of Metasurfaces for Self-Collimated Surface Wave Steering}
%or C-shaped Metasurface for Self-Collimated Surface Wave Steering.
%Spin-dependent Surface Wave Steering by Collimated Metasurfaces}% Force line breaks with \\
%Metasurfaces for Spin-dependent Wave steering
\author{%
Sara M. Kandil\affmark[1], Dia’aaldin J. Bisharat\affmark[2] and Daniel F. Sievenpiper\affmark[1]\\
\affaddr{\affmark[1]\textit{University of California San Diego, Department of Electrical and Computer Engineering \\
 La Jolla, CA 92093, U.S.A}}\\
\affaddr{\affmark[2]\textit{City University of New York, Graduate Center\\ New York, New York 10016, USA}}\\
}
%\author{Sara M. Kandil}
% \altaffiliation[Also at ]{University of California San Diego, Department of Electrical and Computer Engineering}%Lines break automatically or can be forced with \\
%\author{Dia’aaldin Bisharat}%
%\author{Daniel F. Sievenpiper}
%\affiliation{	University of California San Diego, Department of Electrical and Computer Engineering \\  La Jolla, CA 92093, U.S.A }%

%\author{Daniel F. Sievenpiper}
%\affiliation{
%\affiliation{
% \affiliation{%


\date{\today}% It is always \today, today,

\begin{abstract}

Metasurfaces provide unique capability in guiding surface waves and controlling their polarization and dispersion properties. One way to do that is by analyzing their equifrequency contours. Equifrequency contours are the 2D projection of the 3D dispersion diagram. Since they are a k-space map representation of the surface, many of the wave properties can be understood through the Equifrequency contours. In this paper, we investigate numerically and experimentally the engineering of equifrequency contours using C-shape metasurface design. We show the ability to provide high self-collimation as well as spin-dependent wave splitting for the same metasurface by tuning the frequency of operation. We also show the ability to steer the wave along a defined curved path by rotating the C-shape which results in rotating its equifrequency contours. This work demonstrates how engineering equifrequency contours can be used as a powerful tool for controlling the surface wave propagation properties.

%We discuss the symmetry properties of the C-shape that allows it to possess such properties. 

%\begin{description}
%\end{description}
\end{abstract}

\maketitle

%\tableofcontents

%\section{\label{sec:level1}First-level heading:\protect\\ The line
%break was forced \lowercase{via} \textbackslash\textbackslash}
\section{Introduction}

During the last decade, there has been a great interest in making conventional optics such as lenses, waveguides, couplers and polarization-based devices using flat surfaces \cite{yu2014flat,chen2020flat,yu2013flat}. These surface platforms provide the advantage of being scalable and easily integrated on chips. Additionally, they provide additional degrees of freedom for controlling the wave propagation as well as eliminating the accumulated changes in phase and amplitude when propagating over distances as the case in conventional optical systems \cite{capasso2018future}. 

Metasurfaces which are 2D surfaces patterned with subwavelength scatterers are extensively studied as platforms for flat optics. They provide unique capabilities in controlling the dispersion properties, phase and polarization of the propagating wave \cite{chen2016review,li2018metasurfaces}. This can be done by carefully choosing the unit cell design to engineer the interaction between the wave and the surface. Several research has been done to study the wide capabilities of metasurfaces from guiding \cite{quarfoth2013artificial,dia2017guiding,lee2016patterning}, beam focusing \cite{kuznetsov2015planar,ding2020focused}, splitting \cite{ding2020gap,chen2021all,yang2017hyperbolic}, steering \cite{sievenpiper2005forward}, and lensing \cite{liu2020diffractive}. One way to control the interaction between the wave and the surface is by engineering the equifrequency contour of the unit cell design. Equifrequency contour (EFC), also called isofrequency contour (IFC), is the 2D projection of the 3D dispersion diagram at different frequencies \cite{prather2007self, chuang2010complex}. It represents a k-space map of the wave possible trajectories at different frequencies. The direction of the wave is determined by the direction of its group velocity where the group velocity of the wave is defined as $\delta\omega/\delta k$. This can be determined through the EFCs. Anisotropic shapes where some symmetries are broken have interesting, nonconventional EFCs where the contours can vary from elliptical to flat allowing the wave to propagate in one direction (normal to the flat contour) with high self-collimation \cite{witzens2002self}. 

%Here we should have three parts 1- general intro on making conventional optics planar or about collimation and beam steering and all the possible applications. 2- Talk more about self-collimation and EFCs and all different papers for negative refraction, waveguiding,.. 3- Transverse spin and spin-momentum locking 4- Predefined path for wave steering, gradient metasurfaces, what problem will we solve? or how is our results better than others?

Another capability of metasurfaces is that they can control the spin-orbit coupling of surface waves. It was recently shown that surface waves with evanescent tails possess a transverse spin that is locked to the wave momentum. A property termed as Spin-Hall Effect (SHE) where opposite spins propagate in opposite directions \cite{bliokh2012transverse,bliokh2014extraordinary,bliokh2015quantum}. This is analogous to the SHE phenomenon initially discovered in electronic systems \cite{hirsch1999spin}. It is also referred to as spin-momentum locking where spin represents the circular polarization of the electric and/or magnetic field of the surface wave. Several studies showed the ability to achieve spin-dependent unidirectional propagation using metasurfaces with engineered anisotropy \cite{o2014spin,kandil2021c,lin2013polarization}, bandgap materials \cite{bisharat2019electromagnetic,ruan2020analysis}, gradient metasurfaces \cite{huang2013helicity}, and near-field interference with asymmetrically placed dipole sources \cite{picardi2017unidirectional, picardi2018janus,rodriguez2013near}.
\begin{figure}[h!]
	\hspace*{-0.1cm}\includegraphics[width=9cm,height=9cm]{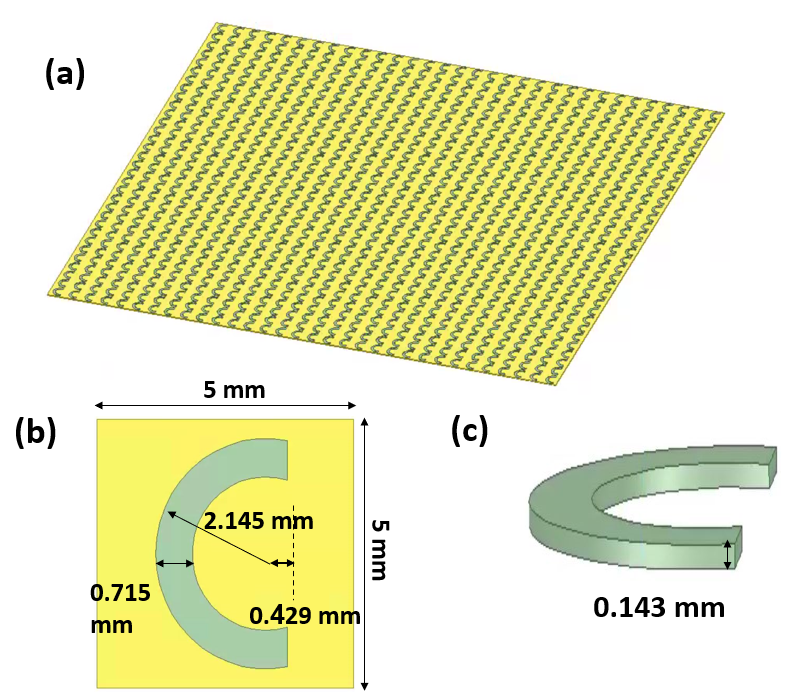}
	\caption{\label{fig:fig1} (a) Schematic showing the C-shape metasurface design. (b) The dimensions of the C-shape unit cell consisting of (d) a C-shape metallic post of 0.143mm thickness placed on a Roger's substrate.} 
\end{figure}
\begin{figure*}
	\hspace*{-0.3cm}\includegraphics[width=19.5cm,height=18cm]{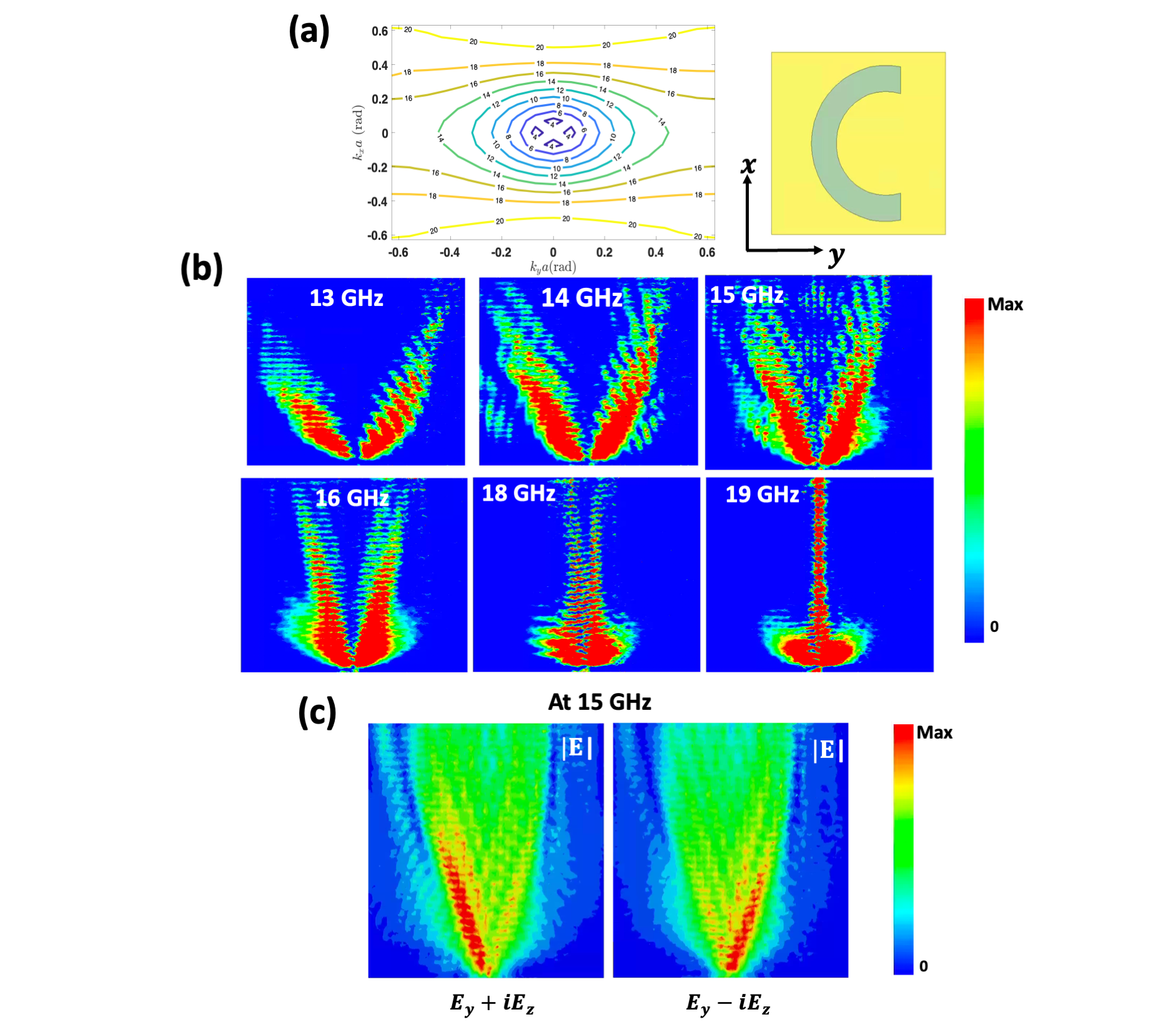}
	\caption{\label{fig:fig2} (a) EFC of the C-shape unit cell. (b) Magnitude of $E_x$ profiles for the C-shape metasurface calculated at frequencies 13GHz to 19GHz when excited with Ey dipole source at bottom center. (c) Magnitude of $E_x$ profiles for the C-shape metasurface calculated at 15GHz when excited at the bottom center with $E_y+iE_z$ (left) and $E_y-iE_z$ (right).} 
\end{figure*}

In this paper, we investigate the different ways the EFC of a metasurface can be engineered to control the surface wave propagation and spin-dependent directionality. We show various wave properties achieved through the same metasurface design such as self-collimation, polarization-based beam splitting and wave steering. The paper is organized as follows: in Section II, we discuss the homogeneous metasurface design formed of metallic C-shape unit cell, its self-collimation and spin-dependent wave propagation. In Section III, we present the inhomogeneous design formed of the same metallic C-shaped unit cells. We discuss how its EFCs change with the rotation angle resulting in surface wave steering along two predefined paths. We study these phenomena using numerical simulations as well as experimental results which are presented in Section IV. The paper is concluded in Section V.
%for higher degree of freedom
%\subsection{\label{sec:level2}Second-level heading: Formatting}
\section{Homogeneous Metasurface}

Fig.\ref{fig:fig1} presents a schematic showing the C-shaped metasurface design we will study throughout this section. As depicted in Fig.\ref{fig:fig1}(b), the unit cell consists of a metallic C-shaped of 0.143mm thickness placed on a Roger's 5880 substrate ($\epsilon_r=2.2$). The C-shape metallic post has a width of 0.715mm and radius of 2.145mm where its edge is extended from the center by 0.429mm. The whole surface is 250mm$\times$250mm. The C-shape metasurface design was studied and optimized numerically using Ansys HFSS. In this section, we will explore different surface wave properties supported by the C-shaped metasurface through studying its equifrequency contours. 

\subsection{Self-Collimation}

Isoropic shapes with small or no asymmetry have EFC close to a circle where the wave propagates equally in all direction. As the asymmetry of the shape increases, the contour becomes flatter with higher wave directionality and hence, higher self-collimation \cite{giden2013broadband}. The C-shape design has broken rotational symmetry along the z- and x-axes which makes it a low-symmetry shape and hence, have high self-collimation. This can be observed from the calculated EFC of the C-shape unit cell shown in Fig.\ref{fig:fig2}(a). It can be shown that the C-shape design possesses different contour shapes which dictate various wave propagation properties. 
At 14 GHz, the EFC is elliptical which then becomes flatter at higher frequencies where high self-collimation takes place. The E-field profiles at different frequency contours are shown in Fig.\ref{fig:fig2}(b) where the surface is excited with an $E_y$ dipole at the bottom center. The magnitude of $E_x$ maps are calculated for each EFC. It can be observed that the wave is split where the split angle decreases with the increase of frequency. At 19 GHz, the surface wave is highly collimated as well as spin-independent. This means that any polarization will excite the wave to propagate with high collimation and zero split angle.

%"I need to talk about the interpretation of the EFC and different properties that can be found" The contour shape corresponds to change from closed elliptical to open, anisotropic to flat 

\subsection{Spin-dependent Propagation}
Fig.\ref{fig:fig2}(c) shows the spin-dependent behavior of the supported surface wave. The EFC calculated for the C-shape shows a spin-based wave splitting property where the wave is split into left-handed and right-handed circularly polarized waves. By excitation of the surface with an $E_y+iE_z$ dipole source, the wave propagates along the left arm where it propagates along the right arm when excited with an $E_y-iE_z$ dipole source.
The spin density is defined as a vector quantity whose direction is normal to the plane of the field circular rotation. The following equation can be used to calculate the spin density of the propagating surface wave \cite{yermakov2016spin}: 
%\begin{figure}[h!]
	
%	\hspace*{-0.1cm}\includegraphics[width=9cm,height=9cm]{Fig3.png}
%	\caption{\label{fig:fig3} (a) Magnitude of $E\textsubscript{z}$ map of the C-shaped chiral waveguide at 22 GHz when excited with an $\bf E\textsubscript{x}+iE\textsubscript{z}$ dipole source (left) } 
%\end{figure}
\begin{figure}
	
	\hspace*{-0.1cm}\includegraphics[width=9cm,height=9cm]{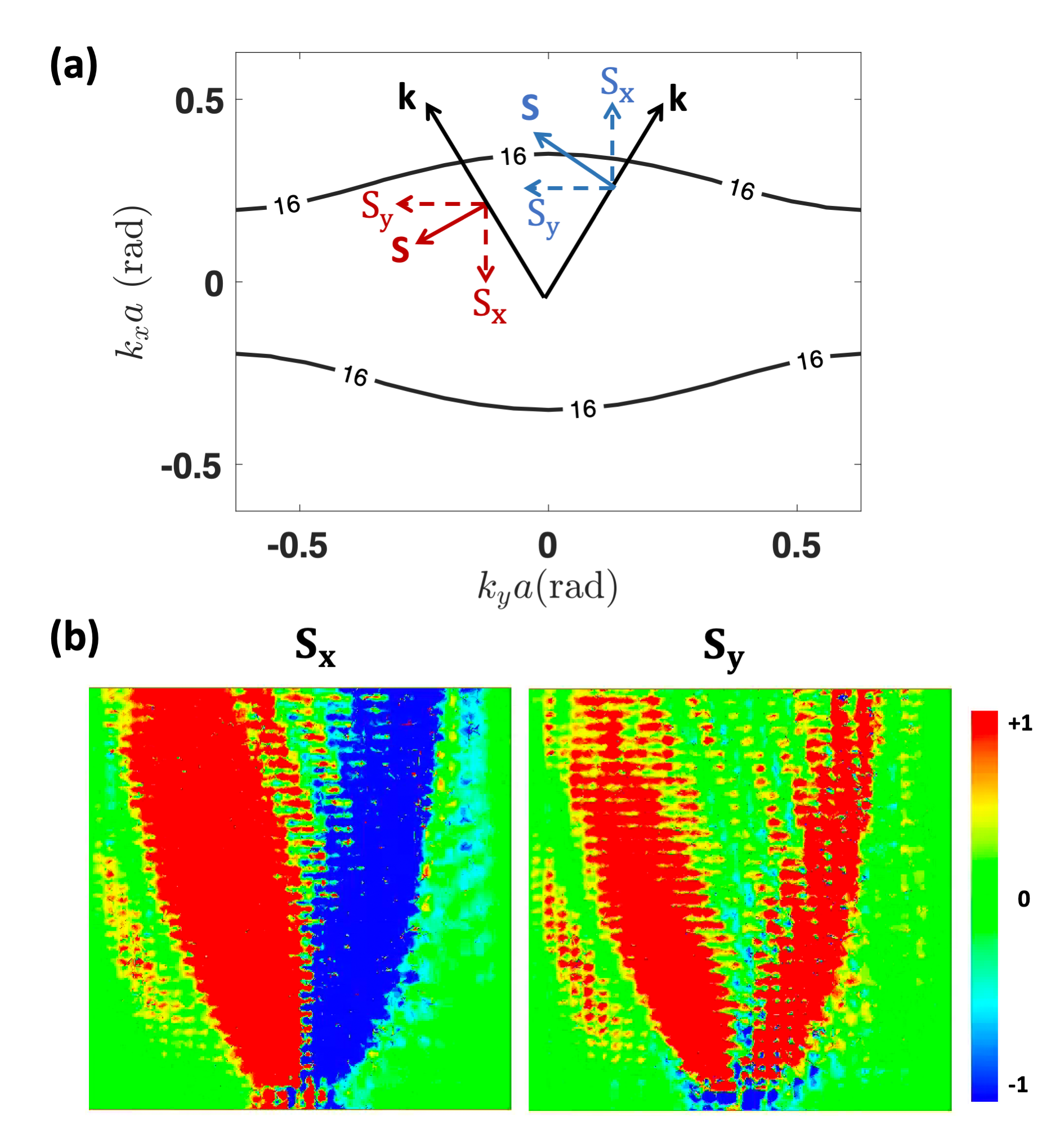}
	\caption{\label{fig:fig3} (a) The 16GHz EFC of the C-shape unit cell where the k-vector directions and transverse spin components are depicted. (b) The x-component (left) and y-component (right) of the transverse spin density vector for the surface wave supported by the C-shape metasurface at 16GHz.} 
\end{figure}

\begin{equation}
\label{eq:one}
\textbf S = \text{Im}\left\{ \frac{\textbf E^*\times \textbf E+ \textbf H^*\times \textbf H}{\big|\textbf E\big|\textsuperscript{2}+\big|\textbf H\big|\textsuperscript{2}} \right\},
\end{equation} where $\textbf{S}$ is the spin density vector in Gaussian units normalized per one photon in units of $\hbar=1$. E and H are the electric and magnetic fields of the surface wave. Fig.\ref{fig:fig3}(a) shows a schematic representation of the two components of the transverse spin with respect to the two wave propagation directions where the $S_x$ component flips sign while the $S_y$ maintains the same sign for both wave propagation directions. Fig.\ref{fig:fig3}(b) shows the numerically calculated spin maps for the two split waves for $S_x$ and $S_y$. 

%As indicated in Fig. 2, by working at a lower frequency contour (yellow contour), the linearly-polarized surface wave will split into two circularly polarized waves with opposite helicities propagating along two different directions at an angle '' from the center axis of the surface.
%A surface wave having two transverse spins can be expressed as follows:
%\begin{eqnarray}
%\textbf E = E_x \textbf a_x+jE_y \textbf a_y-j E_z \textbf a_z,
%\end{eqnarray}
%\begin{equation}
%\textbf H = H_x \textbf a_x-jH_y \textbf a_y+ H_z \textbf a_z, 
%\end{equation}
\section{Inhomogeneous Metasurface}
%talk about gradient metasurfaces
%Engineering the EFC of the unit cell can be used to control the surface wave propagation properties. 
\begin{figure*}
	
	\hspace*{-0.1cm}\includegraphics[width=18cm,height=9cm]{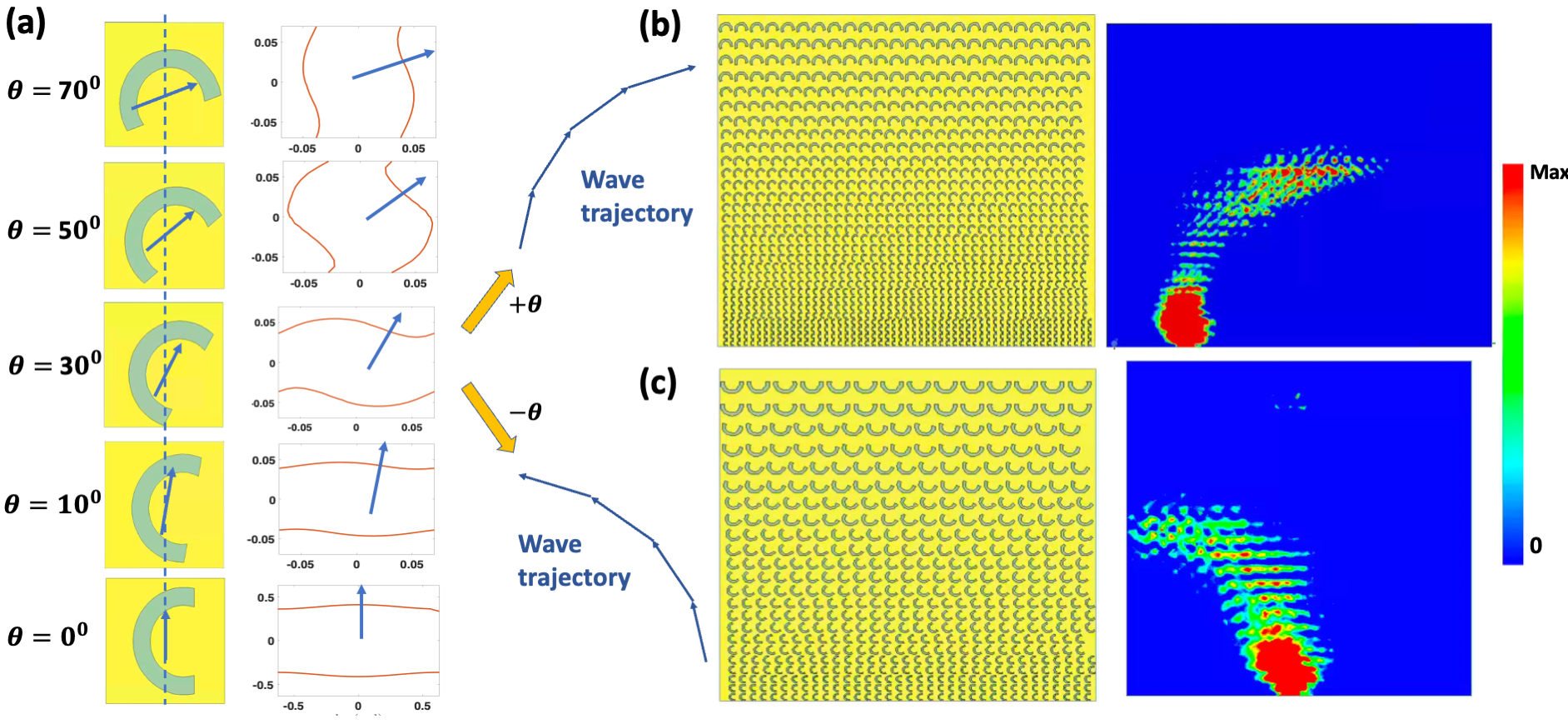}
	\caption{\label{fig:fig4} (a) Schematics showing the rotated C-shape unit cells at rotation angle ($\theta$) from $0^0$ to $70^0$ and their corresponding EFCs. The two wave trajectories formed when $\theta$ is positive and negative are demonstrated.  The designed Inhomogeneous C-shape metasurface designs and their calculated E-field profiles show to steer the surface wave along a curved path towards (b) the right and (c) the left at 19GHz when excited with $E_y-iE_z$.} 
\end{figure*}
%for each angle of rotation of the C-shape unit cell showing the contour rotation as well as the frequency shift.
As demonstrated, the EFC is closely related to the shape of the unit cell; engineering it can be done through different ways. For example, some studies showed that the change of the unit cell from square to rectangular or parallelogram can increase the flatness of the EFCs and hence increase the collimation \cite{gao2008self,xu2008all}. The symmetry of the shape itself can also result in changing the contour shape. For example, as described earlier, a broken rotational symmetry can produce flatter contours along the x- or y-axis. $45^0$ mirror symmetry of a shape can result in tilted contours \cite{kandil2021chiral}. 

In recent decades, there has been a great interest in gradient metasurfaces which are non-periodic surfaces aimed for wavefront manipulation for beam steering applications in near- or far-fields \cite{estakhri2016recent}. Such surfaces can be designed using ray optics approach \cite{yu2011light}, equivalence principle (Huygens metasurfaces) \cite{pfeiffer2013metamaterial}, geometric phase approach \cite{berry1987adiabatic} or using programmable design generation \cite{yang2016programmable}. Here, we present an alternative approach to design such surfaces by engineering EFCs. We show that we are able to steer the surface wave to propagate smoothly along specified curved paths using an inhomogeneous metasurface made of rotated C-shape unit cells. The inhomogeneous design is simply done by mapping the rotated C-shapes and their angle of collimation using the EFCs to form the specified wave path.
%In this section, we show that by rotating the C-shape unit cells and scaling their size, we steer the surface wave to propagate along specified curved paths. 

%By rotating the C-shape unit cell, the EFC rotates as well resulting in rotating the propagation direction of the surface wave in addition to shifting its frequency to a higher value. Hence, in order to route the surface wave along a specified path with optimized impedance matching, we have designed a metasurface consisting of C-shapes that are rotated and scaled-up to route wave direction along a curved path. Fig. 1 shows two metasurface designs for routing an LCP surface wave smoothly along defined curved paths. The electric field is numerically calculated when the surface is excited with an LCP dipole source. It is important to note that the spin studied here is the in-plane transverse spin represented by E-field rotation in the xz axis.

\subsection{Rotation Angle and EFC}

Fig.\ref{fig:fig4}(a) shows the design steps to make the inhomogeneous metasurface designs shown in Fig.\ref{fig:fig4}(b) and (c) through engineering the EFC. With a focus on the C-shape's highly collimated EFC, it can be shown that rotating the C-shape unit cell results in rotating its EFC by almost the same angle of rotation, $\theta$. For simplicity, we show schematics of five rotated unit cells at $\theta=0^0$ to $\theta=70^0$ and their corresponding EFCs. The blue arrows indicate the propagation direction of the wave. The wave trajectory deduced from the EFCs of the rotated C-shapes when using positive $\theta$ values can be shown in Fig.\ref{fig:fig4}(a), top right while the wave trajectory for using negative $\theta$ values where the wave is steered to the left is shown in Fig.\ref{fig:fig4}(a), bottom right.

Fig.\ref{fig:fig4}(b) and (c) show the inhomogeneous metasurface designs for the two wave paths described earlier. The design shown in Fig.\ref{fig:fig4}(b) is composed of 30 columns divided into 9 groups. In each group, $\theta$ is incremented by $10^0$ and the shape is scaled up by a scaling factor, $s$. This is due to the fact that rotating the unit cell results in slight increase in its supported surface wave frequency. To eliminate the frequency mismatch, each rotated C-shape is scaled up by a scaling factor of 1.11. This scaling factor is deduced from mapping the rotation angles and its resulting frequency shift. The scaling factor, $s$, of a unit cell can be defined in terms of its rotation angle as follows:
\begin{equation}
\label{eq:two}
 s = 1.11^{\frac{\theta}{10}}.
\end{equation}

\begin{figure*}[h!]
	
	\hspace*{-0.1cm}\includegraphics[width=18cm,height=18cm]{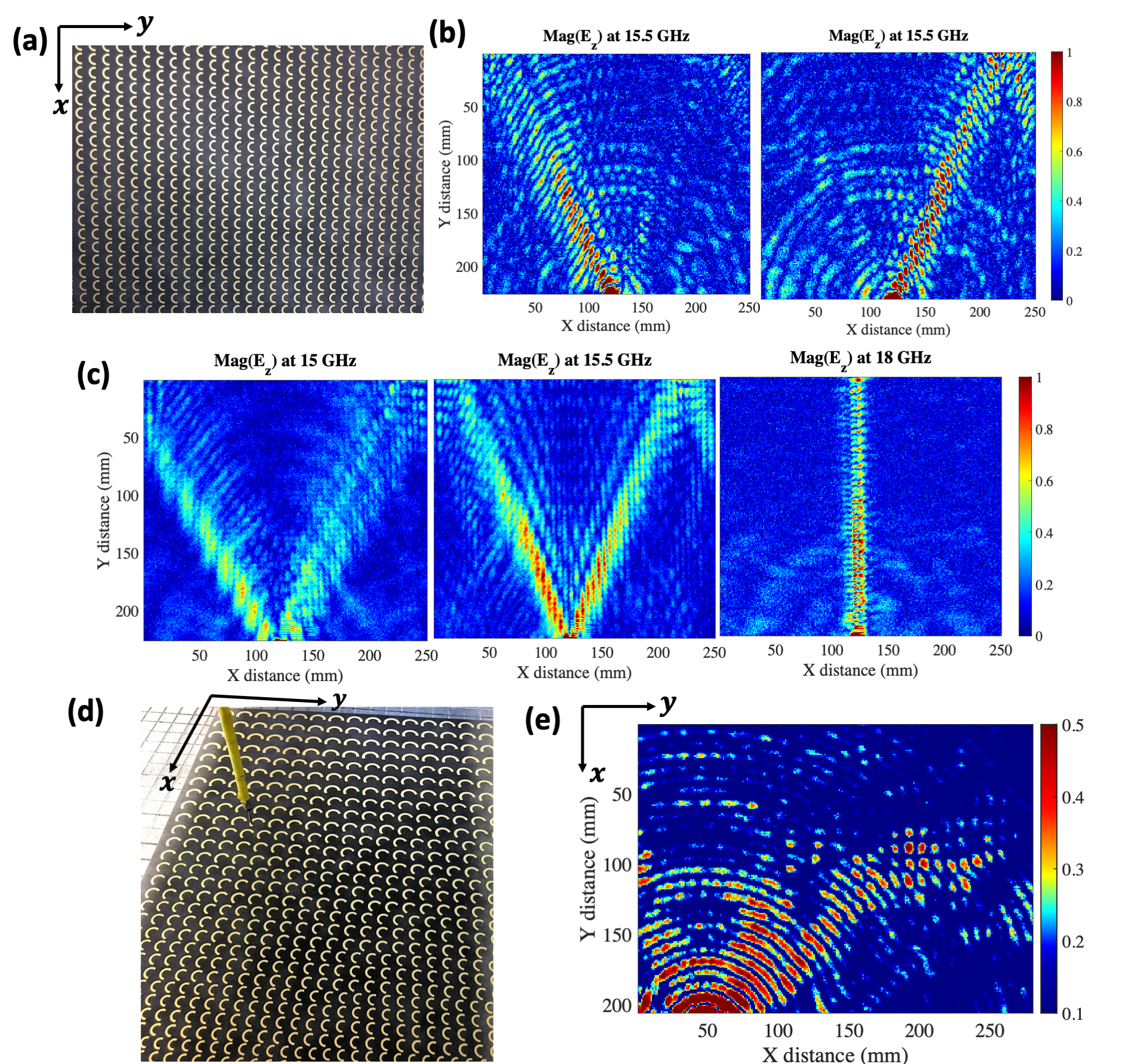}
	\caption{\label{fig:fig5} (a) Photo of the fabricated C-shape metasurface. (b) Measured $E_x$ field maps at frequencies 15GHz to 18GHz showing the different wave split angles when excited with $E_y$ where the highly collimated wave propagation is shown at 18GHz. (c) The measured Ex for the C-shape metasurface at 15.5GHz when excited with $E_y+iE_z$ (left) and $E_y-iE_z$ (right) at the bottom center showing the spin-dependent propagation behavior. (d) Photo of the fabricated inhomogeneous C-shape metasurface. (e) The measured $E_x$ profile at 18GHz showing the surface wave propagation along a curved path.} 
\end{figure*}
\subsection{Surface Wave Steering}
The E-field profiles for the two inhomogeneous metasurface designs at 19GHz are presented on the right in Fig.\ref{fig:fig4}(b) and (c) showing the smooth steering of the surface wave along curved paths. The demonstrated steered wave is spin-independent since we worked with the highly-collimated, spin-independent contour. This means that the same E-field profile can be achieved with a linearly or circularly polarized excitation source. This demonstrates how the EFC engineering for wave steering provides the capability of smoothly steering the surface wave along predetermined paths without much complexity in the designing process where the same unit cell design is used. 
%\begin{figure}[h!]
%The C-shape units cells are scaled up in size as they rotate to minimize the frequency mismatch when the C-shape is rotated
%Fig.\ref{fig:fig4}(b) and (c) show the schematic for two inhomogeneous C-shape metasurfaces for steering the surface wave along a curved path to the right and to the left, respectively. 

%	\hspace*{-0.1cm}\includegraphics[width=9cm,height=9cm]{Fig4_2.PNG}
%	\caption{\label{fig:fig5} (a) Magnitude of $E\textsubscript{z}$ map of the C-shaped chiral waveguide at 22 GHz when excited with an $\bf E\textsubscript{x}+iE\textsubscript{z}$ dipole source (left) } 
%\end{figure}

%It is important to note that the wave steering is optimized for one circular polarization of the wave where the surface is excited with $E_y-iE_z$ dipole source as indicated out in the calculated E-field plots shown on the right
\section{Experimental Results}
The C-shape metasurface was fabricated and measured. Fig.\ref{fig:fig5}(a) shows a photo of the fabricated homogeneous C-shape metasurface while the inhomogeneous surface is shown in Fig.\ref{fig:fig5}(d). The C-shape structures are made of copper which are placed on top of a Roger's 5880 substrate of thickness 0.381mm. An $E_y$ probe used to excite the surface which is placed at the bottom center. Another measuring probe, $E_x$ is attached to a moving station which scans the surface point by point. Both probes are connected to Vector Network Analyzer (VNA) where the magnitude and phase of S21 are measured. The magnitude and phase of the $E_x$ can be extracted for the whole surface at different frequencies showing the different wave propagation properties as shown in Fig.\ref{fig:fig5}(b) which match the simulation results shown in Fig.\ref{fig:fig2}(b). The E-field profile shown on Fig.\ref{fig:fig5}(c) is measured for the homogeneous C-shape surface at 15.5GHz when excited with a circularly polarized (CP) dipole source. The CP dipole excitation is done through excitation with an $E_y$ and $E_z$ dipole sources separately and collecting the magnitude and phase of $E_x$ scan for each excitation. The two scans are then added with a $90^0$ phase shift using the following equation:
\begin{equation}
\textbf{E}_\textsubscript{x}^{y} \pm i\textbf{E}_\textsubscript{x}^{z}= |\textbf{E}_\textsubscript{x}^{y}|\cos\left(\phi_\textsubscript{x}^{y}\right) +|\textbf{E}_\textsubscript{x}^{z}|\cos\left(\phi_\textsubscript{x}^{z}\pm 90^0\right),
\end{equation}
where $\textbf{E}_\textsubscript{x}^{y(z)}$ is the measured $\textbf{E}_\textsubscript{x}$ when excited with a probe along $y (z)$-axis. The $E_z$ profile is measured for the inhomogeneous metasurface at 18.5GHz and shown in Fig.\ref{fig:fig5}(d) which demonstrates the surface wave steering along the predefined path.

%The E-field profile shown on Fig.\ref{fig:fig5}(c) and on Fig.\ref{fig:fig5}(e) are measured for the homogeneous C-shape surface, and inhomogeneous one, respectively, when excited with a circularly polarized (CP) dipole source.

%\subsection{Non-homogeneous Surface}
%\vspace{0.5}
\section{Conclusion}
In this paper we studied numerically and experimentally a metallic C-shape metasurface design and its various wave propagation and polarization properties. We showed that due to the broken rotational symmetry of the C-shape design, it possesses high self-collimation. Additionally, we studied the spin-momentum locking phenomenon in the C-shape by calculating its spin density and showed it is capable of spin-dependent wave splitting. We also showed two inhomogeneous metasurface designs to steer the wave along defined curved paths by rotating the C-shape EFCs and scaling their sizes to eliminate frequency mismatch. This work emphasizes that engineering the EFCs can be simply done to achieve wide wave properties: spin-dependent wave splitting, highly confined waveguiding and wave steering along defined paths. 
\begin{acknowledgments}
This work is supported by the AFOSR under Grant No. FA9550-21-1-0167 as well as the
ONR under Grant No. N00014-20-1-2710.
\end{acknowledgments}

\nocite{*}

\bibliography{apssamp}% Produces the bibliography via BibTeX.

\providecommand{\noopsort}[1]{}\providecommand{\singleletter}[1]{#1}%
\begin{thebibliography}{42}%
\makeatletter
\providecommand \@ifxundefined [1]{%
 \@ifx{#1\undefined}
}%
\providecommand \@ifnum [1]{%
 \ifnum #1\expandafter \@firstoftwo
 \else \expandafter \@secondoftwo
 \fi
}%
\providecommand \@ifx [1]{%
 \ifx #1\expandafter \@firstoftwo
 \else \expandafter \@secondoftwo
 \fi
}%
\providecommand \natexlab [1]{#1}%
\providecommand \enquote  [1]{``#1''}%
\providecommand \bibnamefont  [1]{#1}%
\providecommand \bibfnamefont [1]{#1}%
\providecommand \citenamefont [1]{#1}%
\providecommand \href@noop [0]{\@secondoftwo}%
\providecommand \href [0]{\begingroup \@sanitize@url \@href}%
\providecommand \@href[1]{\@@startlink{#1}\@@href}%
\providecommand \@@href[1]{\endgroup#1\@@endlink}%
\providecommand \@sanitize@url [0]{\catcode `\\12\catcode `\$12\catcode
  `\&12\catcode `\#12\catcode `\^12\catcode `\_12\catcode `\%12\relax}%
\providecommand \@@startlink[1]{}%
\providecommand \@@endlink[0]{}%
\providecommand \url  [0]{\begingroup\@sanitize@url \@url }%
\providecommand \@url [1]{\endgroup\@href {#1}{\urlprefix }}%
\providecommand \urlprefix  [0]{URL }%
\providecommand \Eprint [0]{\href }%
\providecommand \doibase [0]{https://doi.org/}%
\providecommand \selectlanguage [0]{\@gobble}%
\providecommand \bibinfo  [0]{\@secondoftwo}%
\providecommand \bibfield  [0]{\@secondoftwo}%
\providecommand \translation [1]{[#1]}%
\providecommand \BibitemOpen [0]{}%
\providecommand \bibitemStop [0]{}%
\providecommand \bibitemNoStop [0]{.\EOS\space}%
\providecommand \EOS [0]{\spacefactor3000\relax}%
\providecommand \BibitemShut  [1]{\csname bibitem#1\endcsname}%
\let\auto@bib@innerbib\@empty
%</preamble>
\bibitem [{\citenamefont {Yu}\ and\ \citenamefont
  {Capasso}(2014)}]{yu2014flat}%
  \BibitemOpen
  \bibfield  {author} {\bibinfo {author} {\bibfnamefont {N.}~\bibnamefont
  {Yu}}\ and\ \bibinfo {author} {\bibfnamefont {F.}~\bibnamefont {Capasso}},\
  }\bibfield  {title} {\bibinfo {title} {Flat optics with designer
  metasurfaces},\ }\href@noop {} {\bibfield  {journal} {\bibinfo  {journal}
  {Nature materials}\ }\textbf {\bibinfo {volume} {13}},\ \bibinfo {pages}
  {139} (\bibinfo {year} {2014})}\BibitemShut {NoStop}%
\bibitem [{\citenamefont {Chen}\ \emph {et~al.}(2020)\citenamefont {Chen},
  \citenamefont {Zhu},\ and\ \citenamefont {Capasso}}]{chen2020flat}%
  \BibitemOpen
  \bibfield  {author} {\bibinfo {author} {\bibfnamefont {W.~T.}\ \bibnamefont
  {Chen}}, \bibinfo {author} {\bibfnamefont {A.~Y.}\ \bibnamefont {Zhu}},\ and\
  \bibinfo {author} {\bibfnamefont {F.}~\bibnamefont {Capasso}},\ }\bibfield
  {title} {\bibinfo {title} {Flat optics with dispersion-engineered
  metasurfaces},\ }\href@noop {} {\bibfield  {journal} {\bibinfo  {journal}
  {Nature Reviews Materials}\ }\textbf {\bibinfo {volume} {5}},\ \bibinfo
  {pages} {604} (\bibinfo {year} {2020})}\BibitemShut {NoStop}%
\bibitem [{\citenamefont {Yu}\ \emph {et~al.}(2013)\citenamefont {Yu},
  \citenamefont {Genevet}, \citenamefont {Aieta}, \citenamefont {Kats},
  \citenamefont {Blanchard}, \citenamefont {Aoust}, \citenamefont {Tetienne},
  \citenamefont {Gaburro},\ and\ \citenamefont {Capasso}}]{yu2013flat}%
  \BibitemOpen
  \bibfield  {author} {\bibinfo {author} {\bibfnamefont {N.}~\bibnamefont
  {Yu}}, \bibinfo {author} {\bibfnamefont {P.}~\bibnamefont {Genevet}},
  \bibinfo {author} {\bibfnamefont {F.}~\bibnamefont {Aieta}}, \bibinfo
  {author} {\bibfnamefont {M.~A.}\ \bibnamefont {Kats}}, \bibinfo {author}
  {\bibfnamefont {R.}~\bibnamefont {Blanchard}}, \bibinfo {author}
  {\bibfnamefont {G.}~\bibnamefont {Aoust}}, \bibinfo {author} {\bibfnamefont
  {J.-P.}\ \bibnamefont {Tetienne}}, \bibinfo {author} {\bibfnamefont
  {Z.}~\bibnamefont {Gaburro}},\ and\ \bibinfo {author} {\bibfnamefont
  {F.}~\bibnamefont {Capasso}},\ }\bibfield  {title} {\bibinfo {title} {Flat
  optics: controlling wavefronts with optical antenna metasurfaces},\
  }\href@noop {} {\bibfield  {journal} {\bibinfo  {journal} {IEEE Journal of
  Selected Topics in Quantum Electronics}\ }\textbf {\bibinfo {volume} {19}},\
  \bibinfo {pages} {4700423} (\bibinfo {year} {2013})}\BibitemShut {NoStop}%
\bibitem [{\citenamefont {Capasso}(2018)}]{capasso2018future}%
  \BibitemOpen
  \bibfield  {author} {\bibinfo {author} {\bibfnamefont {F.}~\bibnamefont
  {Capasso}},\ }\bibfield  {title} {\bibinfo {title} {The future and promise of
  flat optics: a personal perspective},\ }\href@noop {} {\bibfield  {journal}
  {\bibinfo  {journal} {Nanophotonics}\ }\textbf {\bibinfo {volume} {7}},\
  \bibinfo {pages} {953} (\bibinfo {year} {2018})}\BibitemShut {NoStop}%
\bibitem [{\citenamefont {Chen}\ \emph {et~al.}(2016)\citenamefont {Chen},
  \citenamefont {Taylor},\ and\ \citenamefont {Yu}}]{chen2016review}%
  \BibitemOpen
  \bibfield  {author} {\bibinfo {author} {\bibfnamefont {H.-T.}\ \bibnamefont
  {Chen}}, \bibinfo {author} {\bibfnamefont {A.~J.}\ \bibnamefont {Taylor}},\
  and\ \bibinfo {author} {\bibfnamefont {N.}~\bibnamefont {Yu}},\ }\bibfield
  {title} {\bibinfo {title} {A review of metasurfaces: physics and
  applications},\ }\href@noop {} {\bibfield  {journal} {\bibinfo  {journal}
  {Reports on progress in physics}\ }\textbf {\bibinfo {volume} {79}},\
  \bibinfo {pages} {076401} (\bibinfo {year} {2016})}\BibitemShut {NoStop}%
\bibitem [{\citenamefont {Li}\ \emph {et~al.}(2018)\citenamefont {Li},
  \citenamefont {Singh},\ and\ \citenamefont
  {Sievenpiper}}]{li2018metasurfaces}%
  \BibitemOpen
  \bibfield  {author} {\bibinfo {author} {\bibfnamefont {A.}~\bibnamefont
  {Li}}, \bibinfo {author} {\bibfnamefont {S.}~\bibnamefont {Singh}},\ and\
  \bibinfo {author} {\bibfnamefont {D.}~\bibnamefont {Sievenpiper}},\
  }\bibfield  {title} {\bibinfo {title} {Metasurfaces and their applications},\
  }\href@noop {} {\bibfield  {journal} {\bibinfo  {journal} {Nanophotonics}\
  }\textbf {\bibinfo {volume} {7}},\ \bibinfo {pages} {989} (\bibinfo {year}
  {2018})}\BibitemShut {NoStop}%
\bibitem [{\citenamefont {Quarfoth}\ and\ \citenamefont
  {Sievenpiper}(2013)}]{quarfoth2013artificial}%
  \BibitemOpen
  \bibfield  {author} {\bibinfo {author} {\bibfnamefont {R.}~\bibnamefont
  {Quarfoth}}\ and\ \bibinfo {author} {\bibfnamefont {D.}~\bibnamefont
  {Sievenpiper}},\ }\bibfield  {title} {\bibinfo {title} {Artificial tensor
  impedance surface waveguides},\ }\href@noop {} {\bibfield  {journal}
  {\bibinfo  {journal} {IEEE transactions on antennas and propagation}\
  }\textbf {\bibinfo {volume} {61}},\ \bibinfo {pages} {3597} (\bibinfo {year}
  {2013})}\BibitemShut {NoStop}%
\bibitem [{\citenamefont {Dia’aaldin}\ and\ \citenamefont
  {Sievenpiper}(2017)}]{dia2017guiding}%
  \BibitemOpen
  \bibfield  {author} {\bibinfo {author} {\bibfnamefont {J.~B.}\ \bibnamefont
  {Dia’aaldin}}\ and\ \bibinfo {author} {\bibfnamefont {D.~F.}\ \bibnamefont
  {Sievenpiper}},\ }\bibfield  {title} {\bibinfo {title} {Guiding waves along
  an infinitesimal line between impedance surfaces},\ }\href@noop {} {\bibfield
   {journal} {\bibinfo  {journal} {Physical review letters}\ }\textbf {\bibinfo
  {volume} {119}},\ \bibinfo {pages} {106802} (\bibinfo {year}
  {2017})}\BibitemShut {NoStop}%
\bibitem [{\citenamefont {Lee}\ and\ \citenamefont
  {Sievenpiper}(2016)}]{lee2016patterning}%
  \BibitemOpen
  \bibfield  {author} {\bibinfo {author} {\bibfnamefont {J.}~\bibnamefont
  {Lee}}\ and\ \bibinfo {author} {\bibfnamefont {D.~F.}\ \bibnamefont
  {Sievenpiper}},\ }\bibfield  {title} {\bibinfo {title} {Patterning technique
  for generating arbitrary anisotropic impedance surfaces},\ }\href@noop {}
  {\bibfield  {journal} {\bibinfo  {journal} {IEEE Transactions on Antennas and
  Propagation}\ }\textbf {\bibinfo {volume} {64}},\ \bibinfo {pages} {4725}
  (\bibinfo {year} {2016})}\BibitemShut {NoStop}%
\bibitem [{\citenamefont {Kuznetsov}\ \emph {et~al.}(2015)\citenamefont
  {Kuznetsov}, \citenamefont {Astafev}, \citenamefont {Beruete},\ and\
  \citenamefont {Navarro-C{\'\i}a}}]{kuznetsov2015planar}%
  \BibitemOpen
  \bibfield  {author} {\bibinfo {author} {\bibfnamefont {S.~A.}\ \bibnamefont
  {Kuznetsov}}, \bibinfo {author} {\bibfnamefont {M.~A.}\ \bibnamefont
  {Astafev}}, \bibinfo {author} {\bibfnamefont {M.}~\bibnamefont {Beruete}},\
  and\ \bibinfo {author} {\bibfnamefont {M.}~\bibnamefont {Navarro-C{\'\i}a}},\
  }\bibfield  {title} {\bibinfo {title} {Planar holographic metasurfaces for
  terahertz focusing},\ }\href@noop {} {\bibfield  {journal} {\bibinfo
  {journal} {Scientific reports}\ }\textbf {\bibinfo {volume} {5}},\ \bibinfo
  {pages} {1} (\bibinfo {year} {2015})}\BibitemShut {NoStop}%
\bibitem [{\citenamefont {Ding}\ \emph
  {et~al.}(2020{\natexlab{a}})\citenamefont {Ding}, \citenamefont {Chen},\ and\
  \citenamefont {Bozhevolnyi}}]{ding2020focused}%
  \BibitemOpen
  \bibfield  {author} {\bibinfo {author} {\bibfnamefont {F.}~\bibnamefont
  {Ding}}, \bibinfo {author} {\bibfnamefont {Y.}~\bibnamefont {Chen}},\ and\
  \bibinfo {author} {\bibfnamefont {S.~I.}\ \bibnamefont {Bozhevolnyi}},\
  }\bibfield  {title} {\bibinfo {title} {Focused vortex-beam generation using
  gap-surface plasmon metasurfaces},\ }\href@noop {} {\bibfield  {journal}
  {\bibinfo  {journal} {Nanophotonics}\ }\textbf {\bibinfo {volume} {9}},\
  \bibinfo {pages} {371} (\bibinfo {year} {2020}{\natexlab{a}})}\BibitemShut
  {NoStop}%
\bibitem [{\citenamefont {Ding}\ \emph
  {et~al.}(2020{\natexlab{b}})\citenamefont {Ding}, \citenamefont {Chen},\ and\
  \citenamefont {Bozhevolnyi}}]{ding2020gap}%
  \BibitemOpen
  \bibfield  {author} {\bibinfo {author} {\bibfnamefont {F.}~\bibnamefont
  {Ding}}, \bibinfo {author} {\bibfnamefont {Y.}~\bibnamefont {Chen}},\ and\
  \bibinfo {author} {\bibfnamefont {S.~I.}\ \bibnamefont {Bozhevolnyi}},\
  }\bibfield  {title} {\bibinfo {title} {Gap-surface plasmon metasurfaces for
  linear-polarization conversion, focusing, and beam splitting},\ }\href@noop
  {} {\bibfield  {journal} {\bibinfo  {journal} {Photonics Research}\ }\textbf
  {\bibinfo {volume} {8}},\ \bibinfo {pages} {707} (\bibinfo {year}
  {2020}{\natexlab{b}})}\BibitemShut {NoStop}%
\bibitem [{\citenamefont {Chen}\ \emph {et~al.}(2021)\citenamefont {Chen},
  \citenamefont {Zou}, \citenamefont {Su}, \citenamefont {Tang}, \citenamefont
  {Wang}, \citenamefont {Chen}, \citenamefont {Su},\ and\ \citenamefont
  {Li}}]{chen2021all}%
  \BibitemOpen
  \bibfield  {author} {\bibinfo {author} {\bibfnamefont {X.}~\bibnamefont
  {Chen}}, \bibinfo {author} {\bibfnamefont {H.}~\bibnamefont {Zou}}, \bibinfo
  {author} {\bibfnamefont {M.}~\bibnamefont {Su}}, \bibinfo {author}
  {\bibfnamefont {L.}~\bibnamefont {Tang}}, \bibinfo {author} {\bibfnamefont
  {C.}~\bibnamefont {Wang}}, \bibinfo {author} {\bibfnamefont {S.}~\bibnamefont
  {Chen}}, \bibinfo {author} {\bibfnamefont {C.}~\bibnamefont {Su}},\ and\
  \bibinfo {author} {\bibfnamefont {Y.}~\bibnamefont {Li}},\ }\bibfield
  {title} {\bibinfo {title} {All-dielectric metasurface-based beam splitter
  with arbitrary splitting ratio},\ }\href@noop {} {\bibfield  {journal}
  {\bibinfo  {journal} {Nanomaterials}\ }\textbf {\bibinfo {volume} {11}},\
  \bibinfo {pages} {1137} (\bibinfo {year} {2021})}\BibitemShut {NoStop}%
\bibitem [{\citenamefont {Yang}\ \emph {et~al.}(2017)\citenamefont {Yang},
  \citenamefont {Jing}, \citenamefont {Shen}, \citenamefont {Wang},
  \citenamefont {Zheng}, \citenamefont {Wang}, \citenamefont {Li},
  \citenamefont {Shen}, \citenamefont {Koschny}, \citenamefont {Soukoulis}
  \emph {et~al.}}]{yang2017hyperbolic}%
  \BibitemOpen
  \bibfield  {author} {\bibinfo {author} {\bibfnamefont {Y.}~\bibnamefont
  {Yang}}, \bibinfo {author} {\bibfnamefont {L.}~\bibnamefont {Jing}}, \bibinfo
  {author} {\bibfnamefont {L.}~\bibnamefont {Shen}}, \bibinfo {author}
  {\bibfnamefont {Z.}~\bibnamefont {Wang}}, \bibinfo {author} {\bibfnamefont
  {B.}~\bibnamefont {Zheng}}, \bibinfo {author} {\bibfnamefont
  {H.}~\bibnamefont {Wang}}, \bibinfo {author} {\bibfnamefont {E.}~\bibnamefont
  {Li}}, \bibinfo {author} {\bibfnamefont {N.-H.}\ \bibnamefont {Shen}},
  \bibinfo {author} {\bibfnamefont {T.}~\bibnamefont {Koschny}}, \bibinfo
  {author} {\bibfnamefont {C.~M.}\ \bibnamefont {Soukoulis}}, \emph {et~al.},\
  }\bibfield  {title} {\bibinfo {title} {Hyperbolic spoof plasmonic
  metasurfaces},\ }\href@noop {} {\bibfield  {journal} {\bibinfo  {journal}
  {NPG Asia Materials}\ }\textbf {\bibinfo {volume} {9}},\ \bibinfo {pages}
  {e428} (\bibinfo {year} {2017})}\BibitemShut {NoStop}%
\bibitem [{\citenamefont {Sievenpiper}(2005)}]{sievenpiper2005forward}%
  \BibitemOpen
  \bibfield  {author} {\bibinfo {author} {\bibfnamefont {D.~F.}\ \bibnamefont
  {Sievenpiper}},\ }\bibfield  {title} {\bibinfo {title} {Forward and backward
  leaky wave radiation with large effective aperture from an electronically
  tunable textured surface},\ }\href@noop {} {\bibfield  {journal} {\bibinfo
  {journal} {IEEE transactions on antennas and propagation}\ }\textbf {\bibinfo
  {volume} {53}},\ \bibinfo {pages} {236} (\bibinfo {year} {2005})}\BibitemShut
  {NoStop}%
\bibitem [{\citenamefont {Liu}\ \emph {et~al.}(2020)\citenamefont {Liu},
  \citenamefont {Cheng}, \citenamefont {Tian},\ and\ \citenamefont
  {Chen}}]{liu2020diffractive}%
  \BibitemOpen
  \bibfield  {author} {\bibinfo {author} {\bibfnamefont {W.}~\bibnamefont
  {Liu}}, \bibinfo {author} {\bibfnamefont {H.}~\bibnamefont {Cheng}}, \bibinfo
  {author} {\bibfnamefont {J.}~\bibnamefont {Tian}},\ and\ \bibinfo {author}
  {\bibfnamefont {S.}~\bibnamefont {Chen}},\ }\bibfield  {title} {\bibinfo
  {title} {Diffractive metalens: from fundamentals, practical applications to
  current trends},\ }\href@noop {} {\bibfield  {journal} {\bibinfo  {journal}
  {Advances in Physics: X}\ }\textbf {\bibinfo {volume} {5}},\ \bibinfo {pages}
  {1742584} (\bibinfo {year} {2020})}\BibitemShut {NoStop}%
\bibitem [{\citenamefont {Prather}\ \emph {et~al.}(2007)\citenamefont
  {Prather}, \citenamefont {Shi}, \citenamefont {Murakowski}, \citenamefont
  {Schneider}, \citenamefont {Sharkawy}, \citenamefont {Chen}, \citenamefont
  {Miao},\ and\ \citenamefont {Martin}}]{prather2007self}%
  \BibitemOpen
  \bibfield  {author} {\bibinfo {author} {\bibfnamefont {D.~W.}\ \bibnamefont
  {Prather}}, \bibinfo {author} {\bibfnamefont {S.}~\bibnamefont {Shi}},
  \bibinfo {author} {\bibfnamefont {J.}~\bibnamefont {Murakowski}}, \bibinfo
  {author} {\bibfnamefont {G.~J.}\ \bibnamefont {Schneider}}, \bibinfo {author}
  {\bibfnamefont {A.}~\bibnamefont {Sharkawy}}, \bibinfo {author}
  {\bibfnamefont {C.}~\bibnamefont {Chen}}, \bibinfo {author} {\bibfnamefont
  {B.}~\bibnamefont {Miao}},\ and\ \bibinfo {author} {\bibfnamefont
  {R.}~\bibnamefont {Martin}},\ }\bibfield  {title} {\bibinfo {title}
  {Self-collimation in photonic crystal structures: a new paradigm for
  applications and device development},\ }\href@noop {} {\bibfield  {journal}
  {\bibinfo  {journal} {Journal of Physics D: Applied Physics}\ }\textbf
  {\bibinfo {volume} {40}},\ \bibinfo {pages} {2635} (\bibinfo {year}
  {2007})}\BibitemShut {NoStop}%
\bibitem [{\citenamefont {Chuang}(2010)}]{chuang2010complex}%
  \BibitemOpen
  \bibfield  {author} {\bibinfo {author} {\bibfnamefont {Y.-C.}\ \bibnamefont
  {Chuang}},\ }\emph {\bibinfo {title} {Complex photonic crystals for broadband
  “all-angle” self-collimation}},\ \href@noop {} {Ph.D. thesis},\ \bibinfo
  {school} {The University of North Carolina at Charlotte} (\bibinfo {year}
  {2010})\BibitemShut {NoStop}%
\bibitem [{\citenamefont {Witzens}\ \emph {et~al.}(2002)\citenamefont
  {Witzens}, \citenamefont {Loncar},\ and\ \citenamefont
  {Scherer}}]{witzens2002self}%
  \BibitemOpen
  \bibfield  {author} {\bibinfo {author} {\bibfnamefont {J.}~\bibnamefont
  {Witzens}}, \bibinfo {author} {\bibfnamefont {M.}~\bibnamefont {Loncar}},\
  and\ \bibinfo {author} {\bibfnamefont {A.}~\bibnamefont {Scherer}},\
  }\bibfield  {title} {\bibinfo {title} {Self-collimation in planar photonic
  crystals},\ }\href@noop {} {\bibfield  {journal} {\bibinfo  {journal} {IEEE
  Journal of Selected Topics in Quantum Electronics}\ }\textbf {\bibinfo
  {volume} {8}},\ \bibinfo {pages} {1246} (\bibinfo {year} {2002})}\BibitemShut
  {NoStop}%
\bibitem [{\citenamefont {Bliokh}\ and\ \citenamefont
  {Nori}(2012)}]{bliokh2012transverse}%
  \BibitemOpen
  \bibfield  {author} {\bibinfo {author} {\bibfnamefont {K.~Y.}\ \bibnamefont
  {Bliokh}}\ and\ \bibinfo {author} {\bibfnamefont {F.}~\bibnamefont {Nori}},\
  }\bibfield  {title} {\bibinfo {title} {Transverse spin of a surface
  polariton},\ }\href@noop {} {\bibfield  {journal} {\bibinfo  {journal}
  {Physical review A}\ }\textbf {\bibinfo {volume} {85}},\ \bibinfo {pages}
  {061801} (\bibinfo {year} {2012})}\BibitemShut {NoStop}%
\bibitem [{\citenamefont {Bliokh}\ \emph {et~al.}(2014)\citenamefont {Bliokh},
  \citenamefont {Bekshaev},\ and\ \citenamefont
  {Nori}}]{bliokh2014extraordinary}%
  \BibitemOpen
  \bibfield  {author} {\bibinfo {author} {\bibfnamefont {K.~Y.}\ \bibnamefont
  {Bliokh}}, \bibinfo {author} {\bibfnamefont {A.~Y.}\ \bibnamefont
  {Bekshaev}},\ and\ \bibinfo {author} {\bibfnamefont {F.}~\bibnamefont
  {Nori}},\ }\bibfield  {title} {\bibinfo {title} {Extraordinary momentum and
  spin in evanescent waves},\ }\href@noop {} {\bibfield  {journal} {\bibinfo
  {journal} {Nature communications}\ }\textbf {\bibinfo {volume} {5}},\
  \bibinfo {pages} {1} (\bibinfo {year} {2014})}\BibitemShut {NoStop}%
\bibitem [{\citenamefont {Bliokh}\ \emph {et~al.}(2015)\citenamefont {Bliokh},
  \citenamefont {Smirnova},\ and\ \citenamefont {Nori}}]{bliokh2015quantum}%
  \BibitemOpen
  \bibfield  {author} {\bibinfo {author} {\bibfnamefont {K.~Y.}\ \bibnamefont
  {Bliokh}}, \bibinfo {author} {\bibfnamefont {D.}~\bibnamefont {Smirnova}},\
  and\ \bibinfo {author} {\bibfnamefont {F.}~\bibnamefont {Nori}},\ }\bibfield
  {title} {\bibinfo {title} {Quantum spin hall effect of light},\ }\href@noop
  {} {\bibfield  {journal} {\bibinfo  {journal} {Science}\ }\textbf {\bibinfo
  {volume} {348}},\ \bibinfo {pages} {1448} (\bibinfo {year}
  {2015})}\BibitemShut {NoStop}%
\bibitem [{\citenamefont {Hirsch}(1999)}]{hirsch1999spin}%
  \BibitemOpen
  \bibfield  {author} {\bibinfo {author} {\bibfnamefont {J.}~\bibnamefont
  {Hirsch}},\ }\bibfield  {title} {\bibinfo {title} {Spin hall effect},\
  }\href@noop {} {\bibfield  {journal} {\bibinfo  {journal} {Physical review
  letters}\ }\textbf {\bibinfo {volume} {83}},\ \bibinfo {pages} {1834}
  (\bibinfo {year} {1999})}\BibitemShut {NoStop}%
\bibitem [{\citenamefont {O’connor}\ \emph {et~al.}(2014)\citenamefont
  {O’connor}, \citenamefont {Ginzburg}, \citenamefont
  {Rodr{\'\i}guez-Fortu{\~n}o}, \citenamefont {Wurtz},\ and\ \citenamefont
  {Zayats}}]{o2014spin}%
  \BibitemOpen
  \bibfield  {author} {\bibinfo {author} {\bibfnamefont {D.}~\bibnamefont
  {O’connor}}, \bibinfo {author} {\bibfnamefont {P.}~\bibnamefont
  {Ginzburg}}, \bibinfo {author} {\bibfnamefont {F.~J.}\ \bibnamefont
  {Rodr{\'\i}guez-Fortu{\~n}o}}, \bibinfo {author} {\bibfnamefont {G.~A.}\
  \bibnamefont {Wurtz}},\ and\ \bibinfo {author} {\bibfnamefont {A.~V.}\
  \bibnamefont {Zayats}},\ }\bibfield  {title} {\bibinfo {title} {Spin--orbit
  coupling in surface plasmon scattering by nanostructures},\ }\href@noop {}
  {\bibfield  {journal} {\bibinfo  {journal} {Nature communications}\ }\textbf
  {\bibinfo {volume} {5}},\ \bibinfo {pages} {1} (\bibinfo {year}
  {2014})}\BibitemShut {NoStop}%
\bibitem [{\citenamefont {Kandil}\ and\ \citenamefont
  {Sievenpiper}(2021)}]{kandil2021c}%
  \BibitemOpen
  \bibfield  {author} {\bibinfo {author} {\bibfnamefont {S.~M.}\ \bibnamefont
  {Kandil}}\ and\ \bibinfo {author} {\bibfnamefont {D.~F.}\ \bibnamefont
  {Sievenpiper}},\ }\bibfield  {title} {\bibinfo {title} {C-shaped chiral
  waveguide for spin-dependent unidirectional propagation},\ }\href@noop {}
  {\bibfield  {journal} {\bibinfo  {journal} {Applied Physics Letters}\
  }\textbf {\bibinfo {volume} {118}},\ \bibinfo {pages} {101104} (\bibinfo
  {year} {2021})}\BibitemShut {NoStop}%
\bibitem [{\citenamefont {Lin}\ \emph {et~al.}(2013)\citenamefont {Lin},
  \citenamefont {Mueller}, \citenamefont {Wang}, \citenamefont {Yuan},
  \citenamefont {Antoniou}, \citenamefont {Yuan},\ and\ \citenamefont
  {Capasso}}]{lin2013polarization}%
  \BibitemOpen
  \bibfield  {author} {\bibinfo {author} {\bibfnamefont {J.}~\bibnamefont
  {Lin}}, \bibinfo {author} {\bibfnamefont {J.~B.}\ \bibnamefont {Mueller}},
  \bibinfo {author} {\bibfnamefont {Q.}~\bibnamefont {Wang}}, \bibinfo {author}
  {\bibfnamefont {G.}~\bibnamefont {Yuan}}, \bibinfo {author} {\bibfnamefont
  {N.}~\bibnamefont {Antoniou}}, \bibinfo {author} {\bibfnamefont {X.-C.}\
  \bibnamefont {Yuan}},\ and\ \bibinfo {author} {\bibfnamefont
  {F.}~\bibnamefont {Capasso}},\ }\bibfield  {title} {\bibinfo {title}
  {Polarization-controlled tunable directional coupling of surface plasmon
  polaritons},\ }\href@noop {} {\bibfield  {journal} {\bibinfo  {journal}
  {Science}\ }\textbf {\bibinfo {volume} {340}},\ \bibinfo {pages} {331}
  (\bibinfo {year} {2013})}\BibitemShut {NoStop}%
\bibitem [{\citenamefont {Bisharat}\ and\ \citenamefont
  {Sievenpiper}(2019)}]{bisharat2019electromagnetic}%
  \BibitemOpen
  \bibfield  {author} {\bibinfo {author} {\bibfnamefont {D.~J.}\ \bibnamefont
  {Bisharat}}\ and\ \bibinfo {author} {\bibfnamefont {D.~F.}\ \bibnamefont
  {Sievenpiper}},\ }\bibfield  {title} {\bibinfo {title} {Electromagnetic-dual
  metasurfaces for topological states along a 1d interface},\ }\href@noop {}
  {\bibfield  {journal} {\bibinfo  {journal} {Laser \& Photonics Reviews}\
  }\textbf {\bibinfo {volume} {13}},\ \bibinfo {pages} {1900126} (\bibinfo
  {year} {2019})}\BibitemShut {NoStop}%
\bibitem [{\citenamefont {Ruan}\ \emph {et~al.}(2020)\citenamefont {Ruan},
  \citenamefont {He}, \citenamefont {Zhao},\ and\ \citenamefont
  {Dong}}]{ruan2020analysis}%
  \BibitemOpen
  \bibfield  {author} {\bibinfo {author} {\bibfnamefont {W.-S.}\ \bibnamefont
  {Ruan}}, \bibinfo {author} {\bibfnamefont {X.-T.}\ \bibnamefont {He}},
  \bibinfo {author} {\bibfnamefont {F.-L.}\ \bibnamefont {Zhao}},\ and\
  \bibinfo {author} {\bibfnamefont {J.-W.}\ \bibnamefont {Dong}},\ }\bibfield
  {title} {\bibinfo {title} {Analysis of unidirectional coupling in topological
  valley photonic crystal waveguides},\ }\href@noop {} {\bibfield  {journal}
  {\bibinfo  {journal} {Journal of Lightwave Technology}\ }\textbf {\bibinfo
  {volume} {39}},\ \bibinfo {pages} {889} (\bibinfo {year} {2020})}\BibitemShut
  {NoStop}%
\bibitem [{\citenamefont {Huang}\ \emph {et~al.}(2013)\citenamefont {Huang},
  \citenamefont {Chen}, \citenamefont {Bai}, \citenamefont {Tan}, \citenamefont
  {Jin}, \citenamefont {Zentgraf},\ and\ \citenamefont
  {Zhang}}]{huang2013helicity}%
  \BibitemOpen
  \bibfield  {author} {\bibinfo {author} {\bibfnamefont {L.}~\bibnamefont
  {Huang}}, \bibinfo {author} {\bibfnamefont {X.}~\bibnamefont {Chen}},
  \bibinfo {author} {\bibfnamefont {B.}~\bibnamefont {Bai}}, \bibinfo {author}
  {\bibfnamefont {Q.}~\bibnamefont {Tan}}, \bibinfo {author} {\bibfnamefont
  {G.}~\bibnamefont {Jin}}, \bibinfo {author} {\bibfnamefont {T.}~\bibnamefont
  {Zentgraf}},\ and\ \bibinfo {author} {\bibfnamefont {S.}~\bibnamefont
  {Zhang}},\ }\bibfield  {title} {\bibinfo {title} {Helicity dependent
  directional surface plasmon polariton excitation using a metasurface with
  interfacial phase discontinuity},\ }\href@noop {} {\bibfield  {journal}
  {\bibinfo  {journal} {Light: Science \& Applications}\ }\textbf {\bibinfo
  {volume} {2}},\ \bibinfo {pages} {e70} (\bibinfo {year} {2013})}\BibitemShut
  {NoStop}%
\bibitem [{\citenamefont {Picardi}\ \emph {et~al.}(2017)\citenamefont
  {Picardi}, \citenamefont {Manjavacas}, \citenamefont {Zayats},\ and\
  \citenamefont {Rodr{\'\i}guez-Fortu{\~n}o}}]{picardi2017unidirectional}%
  \BibitemOpen
  \bibfield  {author} {\bibinfo {author} {\bibfnamefont {M.~F.}\ \bibnamefont
  {Picardi}}, \bibinfo {author} {\bibfnamefont {A.}~\bibnamefont {Manjavacas}},
  \bibinfo {author} {\bibfnamefont {A.~V.}\ \bibnamefont {Zayats}},\ and\
  \bibinfo {author} {\bibfnamefont {F.~J.}\ \bibnamefont
  {Rodr{\'\i}guez-Fortu{\~n}o}},\ }\bibfield  {title} {\bibinfo {title}
  {Unidirectional evanescent-wave coupling from circularly polarized electric
  and magnetic dipoles: An angular spectrum approach},\ }\href@noop {}
  {\bibfield  {journal} {\bibinfo  {journal} {Physical Review B}\ }\textbf
  {\bibinfo {volume} {95}},\ \bibinfo {pages} {245416} (\bibinfo {year}
  {2017})}\BibitemShut {NoStop}%
\bibitem [{\citenamefont {Picardi}\ \emph {et~al.}(2018)\citenamefont
  {Picardi}, \citenamefont {Zayats},\ and\ \citenamefont
  {Rodr{\'\i}guez-Fortu{\~n}o}}]{picardi2018janus}%
  \BibitemOpen
  \bibfield  {author} {\bibinfo {author} {\bibfnamefont {M.~F.}\ \bibnamefont
  {Picardi}}, \bibinfo {author} {\bibfnamefont {A.~V.}\ \bibnamefont
  {Zayats}},\ and\ \bibinfo {author} {\bibfnamefont {F.~J.}\ \bibnamefont
  {Rodr{\'\i}guez-Fortu{\~n}o}},\ }\bibfield  {title} {\bibinfo {title} {Janus
  and huygens dipoles: Near-field directionality beyond spin-momentum
  locking},\ }\href@noop {} {\bibfield  {journal} {\bibinfo  {journal}
  {Physical review letters}\ }\textbf {\bibinfo {volume} {120}},\ \bibinfo
  {pages} {117402} (\bibinfo {year} {2018})}\BibitemShut {NoStop}%
\bibitem [{\citenamefont {Rodr{\'\i}guez-Fortu{\~n}o}\ \emph
  {et~al.}(2013)\citenamefont {Rodr{\'\i}guez-Fortu{\~n}o}, \citenamefont
  {Marino}, \citenamefont {Ginzburg}, \citenamefont {O’Connor}, \citenamefont
  {Mart{\'\i}nez}, \citenamefont {Wurtz},\ and\ \citenamefont
  {Zayats}}]{rodriguez2013near}%
  \BibitemOpen
  \bibfield  {author} {\bibinfo {author} {\bibfnamefont {F.~J.}\ \bibnamefont
  {Rodr{\'\i}guez-Fortu{\~n}o}}, \bibinfo {author} {\bibfnamefont
  {G.}~\bibnamefont {Marino}}, \bibinfo {author} {\bibfnamefont
  {P.}~\bibnamefont {Ginzburg}}, \bibinfo {author} {\bibfnamefont
  {D.}~\bibnamefont {O’Connor}}, \bibinfo {author} {\bibfnamefont
  {A.}~\bibnamefont {Mart{\'\i}nez}}, \bibinfo {author} {\bibfnamefont {G.~A.}\
  \bibnamefont {Wurtz}},\ and\ \bibinfo {author} {\bibfnamefont {A.~V.}\
  \bibnamefont {Zayats}},\ }\bibfield  {title} {\bibinfo {title} {Near-field
  interference for the unidirectional excitation of electromagnetic guided
  modes},\ }\href@noop {} {\bibfield  {journal} {\bibinfo  {journal} {Science}\
  }\textbf {\bibinfo {volume} {340}},\ \bibinfo {pages} {328} (\bibinfo {year}
  {2013})}\BibitemShut {NoStop}%
\bibitem [{\citenamefont {Giden}\ \emph {et~al.}(2013)\citenamefont {Giden},
  \citenamefont {Turduev},\ and\ \citenamefont {Kurt}}]{giden2013broadband}%
  \BibitemOpen
  \bibfield  {author} {\bibinfo {author} {\bibfnamefont {I.~H.}\ \bibnamefont
  {Giden}}, \bibinfo {author} {\bibfnamefont {M.}~\bibnamefont {Turduev}},\
  and\ \bibinfo {author} {\bibfnamefont {H.}~\bibnamefont {Kurt}},\ }\bibfield
  {title} {\bibinfo {title} {Broadband super-collimation with low-symmetric
  photonic crystal},\ }\href@noop {} {\bibfield  {journal} {\bibinfo  {journal}
  {Photonics and Nanostructures-Fundamentals and Applications}\ }\textbf
  {\bibinfo {volume} {11}},\ \bibinfo {pages} {132} (\bibinfo {year}
  {2013})}\BibitemShut {NoStop}%
\bibitem [{\citenamefont {Yermakov}\ \emph {et~al.}(2016)\citenamefont
  {Yermakov}, \citenamefont {Ovcharenko}, \citenamefont {Bogdanov},
  \citenamefont {Iorsh}, \citenamefont {Bliokh},\ and\ \citenamefont
  {Kivshar}}]{yermakov2016spin}%
  \BibitemOpen
  \bibfield  {author} {\bibinfo {author} {\bibfnamefont {Y.}~\bibnamefont
  {Yermakov}}, \bibinfo {author} {\bibfnamefont {A.~I.}\ \bibnamefont
  {Ovcharenko}}, \bibinfo {author} {\bibfnamefont {A.~A.}\ \bibnamefont
  {Bogdanov}}, \bibinfo {author} {\bibfnamefont {I.~V.}\ \bibnamefont {Iorsh}},
  \bibinfo {author} {\bibfnamefont {K.~Y.}\ \bibnamefont {Bliokh}},\ and\
  \bibinfo {author} {\bibfnamefont {Y.~S.}\ \bibnamefont {Kivshar}},\
  }\bibfield  {title} {\bibinfo {title} {Spin control of light with hyperbolic
  metasurfaces},\ }\href@noop {} {\bibfield  {journal} {\bibinfo  {journal}
  {Physical Review B}\ }\textbf {\bibinfo {volume} {94}},\ \bibinfo {pages}
  {075446} (\bibinfo {year} {2016})}\BibitemShut {NoStop}%
\bibitem [{\citenamefont {Gao}\ \emph {et~al.}(2008)\citenamefont {Gao},
  \citenamefont {Zhou},\ and\ \citenamefont {Citrin}}]{gao2008self}%
  \BibitemOpen
  \bibfield  {author} {\bibinfo {author} {\bibfnamefont {D.}~\bibnamefont
  {Gao}}, \bibinfo {author} {\bibfnamefont {Z.}~\bibnamefont {Zhou}},\ and\
  \bibinfo {author} {\bibfnamefont {D.~S.}\ \bibnamefont {Citrin}},\ }\bibfield
   {title} {\bibinfo {title} {Self-collimated waveguide bends and partial
  bandgap reflection of photonic crystals with parallelogram lattice},\
  }\href@noop {} {\bibfield  {journal} {\bibinfo  {journal} {JOSA A}\ }\textbf
  {\bibinfo {volume} {25}},\ \bibinfo {pages} {791} (\bibinfo {year}
  {2008})}\BibitemShut {NoStop}%
\bibitem [{\citenamefont {Xu}\ \emph {et~al.}(2008)\citenamefont {Xu},
  \citenamefont {Chen}, \citenamefont {Lan}, \citenamefont {Guo}, \citenamefont
  {Hu},\ and\ \citenamefont {Wu}}]{xu2008all}%
  \BibitemOpen
  \bibfield  {author} {\bibinfo {author} {\bibfnamefont {Y.}~\bibnamefont
  {Xu}}, \bibinfo {author} {\bibfnamefont {X.-J.}\ \bibnamefont {Chen}},
  \bibinfo {author} {\bibfnamefont {S.}~\bibnamefont {Lan}}, \bibinfo {author}
  {\bibfnamefont {Q.}~\bibnamefont {Guo}}, \bibinfo {author} {\bibfnamefont
  {W.}~\bibnamefont {Hu}},\ and\ \bibinfo {author} {\bibfnamefont {L.-J.}\
  \bibnamefont {Wu}},\ }\bibfield  {title} {\bibinfo {title} {The all-angle
  self-collimating phenomenon in photonic crystals with rectangular symmetry},\
  }\href@noop {} {\bibfield  {journal} {\bibinfo  {journal} {Journal of Optics
  A: Pure and Applied Optics}\ }\textbf {\bibinfo {volume} {10}},\ \bibinfo
  {pages} {085201} (\bibinfo {year} {2008})}\BibitemShut {NoStop}%
\bibitem [{\citenamefont {Kandil}\ \emph {et~al.}()\citenamefont {Kandil},
  \citenamefont {Bisharat},\ and\ \citenamefont
  {Sievenpiper}}]{kandil2021chiral}%
  \BibitemOpen
  \bibfield  {author} {\bibinfo {author} {\bibfnamefont {S.~M.}\ \bibnamefont
  {Kandil}}, \bibinfo {author} {\bibfnamefont {D.~J.}\ \bibnamefont
  {Bisharat}},\ and\ \bibinfo {author} {\bibfnamefont {D.~F.}\ \bibnamefont
  {Sievenpiper}},\ }\bibfield  {title} {\bibinfo {title} {Chiral surface wave
  propagation with anomalous spin-momentum locking},\ }\href@noop {} {\bibinfo
  {journal} {ACS Photonics}\ }\BibitemShut {NoStop}%
\bibitem [{\citenamefont {Estakhri}\ and\ \citenamefont
  {Al{\`u}}(2016)}]{estakhri2016recent}%
  \BibitemOpen
\bibfield  {journal} {  }\bibfield  {author} {\bibinfo {author} {\bibfnamefont
  {N.~M.}\ \bibnamefont {Estakhri}}\ and\ \bibinfo {author} {\bibfnamefont
  {A.}~\bibnamefont {Al{\`u}}},\ }\bibfield  {title} {\bibinfo {title} {Recent
  progress in gradient metasurfaces},\ }\href@noop {} {\bibfield  {journal}
  {\bibinfo  {journal} {JOSA B}\ }\textbf {\bibinfo {volume} {33}},\ \bibinfo
  {pages} {A21} (\bibinfo {year} {2016})}\BibitemShut {NoStop}%
\bibitem [{\citenamefont {Yu}\ \emph {et~al.}(2011)\citenamefont {Yu},
  \citenamefont {Genevet}, \citenamefont {Kats}, \citenamefont {Aieta},
  \citenamefont {Tetienne}, \citenamefont {Capasso},\ and\ \citenamefont
  {Gaburro}}]{yu2011light}%
  \BibitemOpen
  \bibfield  {author} {\bibinfo {author} {\bibfnamefont {N.}~\bibnamefont
  {Yu}}, \bibinfo {author} {\bibfnamefont {P.}~\bibnamefont {Genevet}},
  \bibinfo {author} {\bibfnamefont {M.~A.}\ \bibnamefont {Kats}}, \bibinfo
  {author} {\bibfnamefont {F.}~\bibnamefont {Aieta}}, \bibinfo {author}
  {\bibfnamefont {J.-P.}\ \bibnamefont {Tetienne}}, \bibinfo {author}
  {\bibfnamefont {F.}~\bibnamefont {Capasso}},\ and\ \bibinfo {author}
  {\bibfnamefont {Z.}~\bibnamefont {Gaburro}},\ }\bibfield  {title} {\bibinfo
  {title} {Light propagation with phase discontinuities: generalized laws of
  reflection and refraction},\ }\href@noop {} {\bibfield  {journal} {\bibinfo
  {journal} {science}\ }\textbf {\bibinfo {volume} {334}},\ \bibinfo {pages}
  {333} (\bibinfo {year} {2011})}\BibitemShut {NoStop}%
\bibitem [{\citenamefont {Pfeiffer}\ and\ \citenamefont
  {Grbic}(2013)}]{pfeiffer2013metamaterial}%
  \BibitemOpen
  \bibfield  {author} {\bibinfo {author} {\bibfnamefont {C.}~\bibnamefont
  {Pfeiffer}}\ and\ \bibinfo {author} {\bibfnamefont {A.}~\bibnamefont
  {Grbic}},\ }\bibfield  {title} {\bibinfo {title} {Metamaterial huygens’
  surfaces: tailoring wave fronts with reflectionless sheets},\ }\href@noop {}
  {\bibfield  {journal} {\bibinfo  {journal} {Physical review letters}\
  }\textbf {\bibinfo {volume} {110}},\ \bibinfo {pages} {197401} (\bibinfo
  {year} {2013})}\BibitemShut {NoStop}%
\bibitem [{\citenamefont {Berry}(1987)}]{berry1987adiabatic}%
  \BibitemOpen
  \bibfield  {author} {\bibinfo {author} {\bibfnamefont {M.~V.}\ \bibnamefont
  {Berry}},\ }\bibfield  {title} {\bibinfo {title} {The adiabatic phase and
  pancharatnam's phase for polarized light},\ }\href@noop {} {\bibfield
  {journal} {\bibinfo  {journal} {Journal of Modern Optics}\ }\textbf {\bibinfo
  {volume} {34}},\ \bibinfo {pages} {1401} (\bibinfo {year}
  {1987})}\BibitemShut {NoStop}%
\bibitem [{\citenamefont {Yang}\ \emph {et~al.}(2016)\citenamefont {Yang},
  \citenamefont {Cao}, \citenamefont {Yang}, \citenamefont {Gao}, \citenamefont
  {Xu}, \citenamefont {Li}, \citenamefont {Chen}, \citenamefont {Zhao},
  \citenamefont {Zheng},\ and\ \citenamefont {Li}}]{yang2016programmable}%
  \BibitemOpen
  \bibfield  {author} {\bibinfo {author} {\bibfnamefont {H.}~\bibnamefont
  {Yang}}, \bibinfo {author} {\bibfnamefont {X.}~\bibnamefont {Cao}}, \bibinfo
  {author} {\bibfnamefont {F.}~\bibnamefont {Yang}}, \bibinfo {author}
  {\bibfnamefont {J.}~\bibnamefont {Gao}}, \bibinfo {author} {\bibfnamefont
  {S.}~\bibnamefont {Xu}}, \bibinfo {author} {\bibfnamefont {M.}~\bibnamefont
  {Li}}, \bibinfo {author} {\bibfnamefont {X.}~\bibnamefont {Chen}}, \bibinfo
  {author} {\bibfnamefont {Y.}~\bibnamefont {Zhao}}, \bibinfo {author}
  {\bibfnamefont {Y.}~\bibnamefont {Zheng}},\ and\ \bibinfo {author}
  {\bibfnamefont {S.}~\bibnamefont {Li}},\ }\bibfield  {title} {\bibinfo
  {title} {A programmable metasurface with dynamic polarization, scattering and
  focusing control},\ }\href@noop {} {\bibfield  {journal} {\bibinfo  {journal}
  {Scientific reports}\ }\textbf {\bibinfo {volume} {6}},\ \bibinfo {pages} {1}
  (\bibinfo {year} {2016})}\BibitemShut {NoStop}%
\end{thebibliography}%


\providecommand{\noopsort}[1]{}\providecommand{\singleletter}[1]{#1}%
%

\end{document}